\magnification=\magstep1
\baselineskip=16pt
\hfuzz=6pt

\centerline{\bf Unconventional Quantum Computing Devices}

\bigskip

\centerline{Seth Lloyd}

\centerline{Mechanical Engineering}

\centerline{MIT 3-160}

\centerline{Cambridge, Mass. 02139}

\bigskip\noindent{\it Abstract:} This paper investigates a 
variety of unconventional quantum computation devices, including
fermionic quantum computers and computers that exploit
nonlinear quantum mechanics.  It is shown that unconventional
quantum computing devices can in principle compute some
quantities more rapidly than `conventional' quantum
computers.

\bigskip

Computers are physical: what they can and cannot do is determined
by the laws of physics.  When scientific progress augments
or revises those laws, our picture of what computers can do
changes.  Currently, quantum mechanics is generally accepted
as the fundamental dynamical theory of how physical systems
behave.  Quantum computers can in principle exploit quantum
coherence to perform computational tasks that classical
computers cannot [1-21].  If someday quantum mechanics should turn
out to be incomplete or faulty, then our picture of what
computers can do will change.  In addition, the set of known
quantum phenomena is constantly increasing: essentially any
coherent quantum phenomenon involving nonlinear
interactions between quantum degrees of freedom can in
principle be exploited to perform quantum logic.  This paper
discusses how the revision of fundamental laws and the
discovery of new quantum phenomena can lead to new
technologies and algorithms for quantum computers.

Since new quantum effects are discovered seemingly every
day, let's first discuss two basic tests that a phenomenon
must pass to be able to function as a basis for quantum
computation.  These are 1) The phenomenon must be nonlinear,
and 2) It must be coherent.  To support quantum logic, the
phenomenon must involve some form of nonlinearity, e.g., a
nonlinear interaction between quantum degrees of freedom.
Without such a nonlinearity quantum devices, like linear
classical devices, cannot perform even so simple a nonlinear
operation as an AND gate.  Quantum coherence is a
prerequisite for performing tasks such as factoring using
Shor's algorithm [10], quantum simulation a la Feynman [11] and
Lloyd [12], or Grover's data-base search algorithm [13], all of
which require extended manipulations of coherent quantum
superpositions.  

The requirements of nonlinearity and coherence are not only
necessary for a phenomenon to support quantum computation,
they are also in principle sufficient.  As shown in [14-15], essentially
any nonlinear interaction between quantum degrees of freedom
suffices to construct universal quantum logic gates that can
be assembled into a quantum computer.  In addition, the work
of Preskill {\it et al.} [18] on robust quantum computation
shows that an error rate of no more than $10^{-4}$ per quantum logic
operation allows one to perform arbitrarily long quantum
computations in principle.

In practice, of course, few if any quantum phenomena are
likely to prove sufficiently controllable to provide
extended quantum computation.  Promising devices under
current experimental investigation include ion traps [5,7],
high finesse cavities for manipulating light and atoms using
quantum electrodynamics [6], and molecular systems that can be
made to compute using nuclear magnetic resonance [8-9].
Such devices store quantum information on the states of
quantum systems such as photons, atoms, or nuclei, and
accomplish quantum logic by manipulating the interactions
between the systems via the application of semiclassical
potentials such as microwave or laser fields.  We will call
such devices `conventional' quantum computers, if only
because such devices have actually been constructed.

There is another sense in which such computers are
conventional: although the devices described above have
already been used to explore new regimes in physics and to
create and investigate the properties of new and exotic
quantum states of matter, they function according to
well established and well understood laws of physics.
Perhaps the most striking examples of the `conventionality'
of current quantum logic devices are NMR quantum
microprocessors that are operated using techniques that have
been refined for almost half a century.
Ion-trap and quantum electrodynamic quantum computers,
though certainly cutting edge devices, operate in a 
quantum electrodynamic regime where the fundamental
physics has been understood for decades (that is not to
say that new and unexpected physics does not arise
frequently in this regime, rather that there is general agreement
on how to model the dynamics of such devices).

Make no mistake about it: a conventional quantum logic
device is the best kind of quantum logic device to have
around.  It is exactly {\it because} the physics of nuclear
magnetic resonance and quantum electrodynamics are well
understood that devices based on this physics can be used
systematically to construct and manipulate the exotic
quantum states that form the basis for quantum computation.
With that recognition, let us turn to `unconventional'
quantum computers.

Perhaps the most obvious basis for an unconventional quantum
computer is the use of particles with non-Boltzmann
statistics in a refime where these statistics play a key
role in the dynamics of the device.  For example, Lloyd
[16] has proposed the use of fermions as the fundamental
carriers of quantum information, so that a site or state
occupied by a fermion represents a 1 and an unoccupied site
or state represents a 0.  It is straightforward to design a
universal quantum computer using a conditional hopping
dynamics on an array of sites, in which a fermion hops from
one site to another if only if other sites are occupied.

If the array is one-dimensional, then such a fermionic
quantum computer is equivalent to a conventional quantum
computer via the well-known technique of bosonization.  If
the array is two or more dimensional, however, a local
operation involving fermions on the lattice cannot be mocked
up by a local operation on a conventional quantum computer,
which must explicitly keep track of the phases induced by Fermi
statistics.  As a result, such a fermionic computer can
perform certain operations more rapidly than a conventional
quantum computer.  An obvious example of a problem that can
be solved more rapidly on a fermionic quantum computer is
the problem of simulating a lattice fermionic system in two or more
dimensions.  To get the antisymmetrization right in second
quantized form, a conventional `Boltzmann' quantum computer
takes time proportional to $T\ell^{d-1}$ where $T$ is the time over which
the simulation is to take place, $\ell$ is the length of the
lattice and $d$ is the dimension, while a fermionic quantum
computer takes time proportional to $T$.  (Here we assume
that the computations for both conventional and Fermionic
quantum computers can take advantage of the intrinsic
parallelizability of such simulations: if the computations
are performed serially an additional factro of $\ell^d$ is
required for both types of computer to update each site
sequentially.)

As the lattice size $\ell$ and the dimension $d$ grow large,
the difference between the two types of computer also grows
large.  Indeed, the problem of simulating fermions hopping
on a hypercube of dimension $d$ as $d\rightarrow\infty$ is
evidently exponentially harder on a conventional quantum
computer than a Fermionic quantum computer.  Since a variety
of difficult problems such as the travelling-salesman
problem and data-base search problem can be mapped to
particles hopping on a hypercube, it is interesting to
speculate whether fermionic computers might provide an
exponential speed-up on problems of interest in addition
to quantum simulation.  No such problems are currently
known, however.  Fermionic computers could be realized in
principle by manipulating the ways in which electrons and
holes hop from site to site on a semiconductor lattice
(though problems of decoherence are likely to be relatively
severe for such systems).

It might also be possible to construct bosonic computers
using photons, phonons, or atoms in a Bose-Einstein condensate.
Such systems can be highly coherent and support nonlinear interactions:
phonons and photons can interact in a nonlinear fshion via
their common nonlinear interaction with matter, and atoms in 
a Bose condensate can be made to interact bia quantum 
electrodynamics (by introduction of a cavity) or by collisions.
So far, however, the feature of Bose condensates that makes
them so interesting from the point of view of physics ---
all particles in the same state --- makes them less
interesting from the point of view of quantum computation.
Many particles in the same state, which can be manipulated
coherently by a variety of techniques, explore the same
volume of Hilbert space as a single particle in that state.
As a result, it is unclear how such a bosonic system could
provide a speed-up over conventional quantum computation.
More promising than Bose condensates from the perspective
of quantum computation and quantum communications, is the
use of cavity quantum electrodynamics to `dial up' or
synthesize arbitrary states of the cavity field.  Such
a use of bosonic states is important for the field of
quantum communications, which requires the ability to
create and manipulate entangled states of the
electromagnetic field.

A third unconventional design for a quantum computer relies
on `exotic' statistics that are neither fermionic nor
bosonic.  Kitaev has recently proposed a quantum computer
architecture based on `anyons,' particles that when
exchanged acquuire an arbitrary phase.  Examples of anyons
include two-dimensional topological defects in lattice
systems of spins with various symmetries.  Kitaev noted
that such anyons could perform quantum logic via
Aharonov-Bohm type interactions [19].  Preskill {\it et al.}
have shown explicitly how anyonic systems could compute in
principle [20], and Lloyd {\it et al.} have proposed methods of
realizing anyons using superconducting circuits (they could
also in principle be constructed using NMR quantum
computers to mock up the anyonic dynamics in an effectively
two-dimensional space of spins) [21].  The advantage of using
anyons for quantum computation is that their nonlocal
topological nature can make them intrinsically
error-correcting and virtually immune to the effects of
noise and interference.

As the technologies of the microscale become better
developed, more and more potential designs for quantum
computers, both conventional and unconventional, are likely
to arise.  Additional technologies that could prove useful
for the construction of quantum logic devices include
photonic crystals, optical hole-burning techniques,
electron spin resonance, quantum dots, superconducting
circuits in the quantum regime, etc.  Since every quantum
degree of freedom can in principle participate in a
computation one cannot {\it a priori} rule out the
possibility of using currently hard to control degrees of
freedom such as quark and gluon in complex nuclei to
process information.  Needless to say, most if not all of
the designs inspired by these technologies are likely to
fail.  There is room for optimism that some such quantum
computer designs will prove practicable, however.

The preceding unconventional designs for quantum computers
were based on existing, experimentally confirmed physical
phenomena (except in the case of non-abelian anyons).  Let
us now turn to designs based on speculative, hypothetical,
and not yet verified phenomena.  (One of the most
interesting of these phenomena is large-scale quantum
computation itself: can we create and systematically
transform entangled states involving hundreds or thousands
of quantum variables?)  A particularly powerful hypothesis
from the point of view of quantum computation is that of nonlinear
quantum mechanics.

The conventional picture of quantum mechanics is that it is
linear in the sense that the superposition principle is
obeyed exactly.  (Of course, quantum systems can still
exhibit nonlinear interactions between degrees of freedom
while continuing to obey the superposition principle.)
Experiment confirms that the superposition principle is
indeed obeyed to a high degree of accuracy.  Nonetheless,
a number of scientists including Weinberg have proposed
nonlinear versions of quantum mechanics in which the
superposition principle is violated.  Many of these
proposals exhibit pathologies such as violations of the
second law of thermodynamics or the capacity for
superluminal communication.  Despite such theoretical
difficulties, it is still possible that quantum mechanics
does indeed possess a small nonlinearity,  even if it currently
seems unlikely.  If a nonlinear operation such as that
proposed by Weinberg can be incorporated in a quantum
logic operation, then the consequences are striking:
NP-complete problems can be solved easily in polynomial
time [17].  Indeed, NP-oracle problems and all problems in
$\#P$ can be solved in polynomial time on such a nonlinear
quantum computer.  

A general proof of this result is given in [17], however, a
simple argument for why this is so can be seen as follows.
Suppose that it is possible to perform a non-unitary
operation on a single qubit that has a positive Lyapunov
exponent over some region: i.e., somewhere on the unit
sphere there exists a line of finite extent along which
application of the operation causes nearby points to move
apart exponentially at a rate $e^{\lambda\Delta\theta}$
proportional to their original angular separation
$\delta\theta$.  Now consider a function $f(x)$ from
$N$ bits to one bit.  We wish to determine whether or not
there exists an $x$ such that $f(x)=1$, and if so, how many
such $x$'s there are.  Using the nonlinear operation with
positive Lyapunov exponent, it is straightforward to construct a
mapping leaves a point on the exponentially expanding
line (call this point $|0\rangle$)
fixed if their are no solutions to the equation $f(x)=1$,
and that maps the point to a nearby point ${\rm cos}(n/2^N)
|0\rangle +{\rm sin}(n/2^N)|1\rangle$ 
along the line if there are exactly $n$
solutions to the equation $f(x)=1$.  Repeated application
of the nonlinear map can be used to drive the points apart
at an exponentional rate: eventually, at a time determined
by the number of qubits $N$, the number of solutions $n$,
and the rate of spreading $\lambda$, the two points will
become macroscopically distinguishable, 
allowing one to determine whether or not
there is a solution and if there is, how many solutions
there are.  The map $f$ need only be applied once, and the
amount of time it takes to reveal the number of solutions
is proportional to $N$.

The fact that nonlinear quantum mechanics allows the
straightforward solution of NP-complete and $\#P$ problems
should probably be regarded as yet another strike against
nonlinear quantum mechanics.  Whether or not quantum
mechanics is linear is a question to be resolved
experimentally, however.  In the unlikely event that
quantum mechanics does turn out to be nonlinear, all our
problems may be solved.

Finally, let us turn our attention to hypothetical 
quantum Theories of Everything, such as string theory.
Such a theory must clearly support quantum computation
since it supports cavity quantum electrodynamics and
nuclear magnetic resonance.  The obvious question to ask is
then, does a Theory of Everything need to support anything
{\it more} than quantum computation?  So far as
experimental evidence is concerned the answer to this
question is apparently No: we have no evident reason to
doubt that the universe is at bottom anything more than a
giant, parallel, quantum information processing machine,
and that the phenomena that we observe and attempt to
characterize are simply outputs of this machine's ongoing
computation.  Of course, just how the universe is carrying
out this computation is likely to remain a question of
great interest for some time.

To summarize: Computers are physical systems, and what they
can do in practice and in principle is circumscribed by the
laws of physics.  The laws of physics in turn permit a wide
variety of quantum computational devices including some
based on nonconventional statistics and exotic effects.  
Modifications made to
the laws of physics have the consequence that what can be
computed in practice and in principle changes.  A
particularly intriguing variation on conventional physics
is nonlinear quantum mechanics which, if true, would allow
hard problems to be solved easily.

\vfill\eject
\noindent {\bf References}
\bigskip

\noindent 1. P. Benioff, `Quantum Mechanical Models of Turing
Machines that Dissipate No Energy,' {\it Physical Review
Letters}, Vol. {\bf 48}, No. 23, pp. 1581-1585 (1982)

\noindent 2. D. Deutsch, `Quantum Theory, the Church-Turing
Principle
and the Universal Quantum Computer,' {\it Proceedings of
the Royal Society of London, A}, Vol. {\bf 400}, pp. 97-117
(1985).

\noindent 3. R.P. Feynman, `Quantum Mechanical Computers,'
{\it Optics
News}, Vol. {\bf 11}, pp. 11-20 (1985); also in {\it
Foundations of Physics}, Vol. {\bf 16}, pp. 507-531 (1986).

\noindent 4. S. Lloyd, `A Potentially Realizable Quantum
Computer,'
{\it Science}, Vol. {\bf 261}, pp. 1569-1571 (1993).

\noindent 5. J.I. Cirac and P. Zoller, `Quantum Computations
with Cold Trapped Ions,' {\it Physical Review Letters},
Vol. {\bf 74}, pp. 4091-4094 (1995).

\noindent 6. Q.A. Turchette, C.J. Hood, W. Lange, H. Mabuchi,
H.J. Kimble, `Measurement of Conditional Phase Shifts
for Quantum Logic,' {\it Physical Review Letters}, Vol.
{\bf 75}, pp. 4710-4713 (1995).

\noindent 7. C. Monroe, D.M. Meekhof, B.E. King, W.M. Itano,
D.J. Wineland, `Demonstration of a Fundamental Quantum Logic Gate,'
{\it Physical Review Letters}, Vol. {\bf 75},
pp. 4714-4717 (1995).

\noindent 8. D.G. Cory, A.F. Fahmy, T.F. Havel, `Nuclear Magnetic
Resonance Spectroscopy: an experimentally accessible paradigm
for quantum computing,' in {\it PhysComp96}, Proceedings of
the Fourth Workshop on Physics and Computation, T. Toffoli, M.
Biafore, J. Le\~ao, eds., New England Complex Systems Institute,
1996, pp. 87-91.

\noindent 9. N.A. Gershenfeld and I.L. Chuang, `Bulk
Spin-Resonance Quantum Computation,' {\it Science}, Vol. {\bf 275}, pp.
350-356 (1997).

\noindent 10. P. Shor, `Algorithms for Quantum Computation:
Discrete Log and Factoring,'
in {\it Proceedings of the 35th Annual Symposium on Foundations
of Computer Science}, S. Goldwasser, Ed., IEEE Computer
Society, Los Alamitos, CA, 1994, pp. 124-134.

\noindent 11. R.P. Feynman, `Simulating Physics with Computers,'
{\it International Journal of Theoretical Physics},
Vol. {\bf 21}, pp. 467-488 (1982).

\noindent 12. S. Lloyd, `Universal Quantum Simulators,' {\it
Science}, Vol. {\bf 273}, pp. 1073-1078 (1996).

\noindent 13. L.K. Grover, `Quantum Mechanics Helps in Searching
for a Needle in a Haystack,' {\it Physical Review Letters},
Vol. {\bf 79}, pp. 325-328 (1997).

\noindent 14. D. Deutsch, A. Barenco, A. Ekert, `Universality
in Quantum Computation,' {\it Proceedings of the Royal
Society of London A}, Vol. {\bf 449}, pp. 669-677 (1995).

\noindent 15. S. Lloyd, `Almost Any Quantum Logic Gate is
Universal,' {\it Physical Review Letters}, Vol. {\bf 75},
pp. 346-349 (1995).

\noindent 16. S. Lloyd, `Fermionic Quantum Computers,'
talk delivered at the Santa Barbara workshop on Physics
of Information, November 1996.

\noindent 17. D. Abrams and S. Lloyd, to be published.

\noindent 18. J. Preskill {\it et al.}, to be published.

\noindent 19. Yu. Kitaev, to be published.

\noindent 20. J. Preskill {\it et al.}, to be published.

\noindent 21. S. Lloyd {\it et al.} to be published.

\vfill\eject\end